\documentclass{osa-article}

\journal{oe}

\articletype{Research Article}

\usepackage{lineno}
\usepackage{color}
\nolinenumbers

\begin{document}

\title{Nanotips for 0.5THz scattering scanning near field microscopy }

\author{Zeliang Zhang\authormark{1,2}, Pengfei Qi\authormark{1,2}, Olga Kosavera\authormark{1,4}, Cheng Gong\authormark{1,2*}, Lie Lin\authormark{1,3}, and Weiwei Liu\authormark{1,2}}

\address{\authormark{1}Institute of Modern Optics, Nankai University, Tianjin 300350, China\\
\authormark{2}Tianjin Key Laboratory of Micro-scale Optical Information Science and Technology, Tianjin 300350, China\\
\authormark{3}Tianjin Key Laboratory of Optoelectronic Sensor and Sensing Network Technology, Tianjin 300350, China
\authormark{4}Faculty of Physics, Lomonosov Moscow State University, Leninskie Gory, Moscow 119991, Russia
}

\email{\authormark{*}gongcheng@nankai.edu.cn} 



\begin{abstract}
This manuscript demonstrates the theory, design, and simulation of scattering scanning near field microscopy (s-SNOM) in 0.5THz. A comprehensive simulation model of nanotips’ geometry, sample materials, and incident field is established to significantly improve the scattering efficiency and spatial resolution to achieve optimal performance. The theoretical model is based on full-wave simulation and dipole moment analysis which can describe the overall nanotip’s geometry information to screen the optimal parameters corresponding to the 0.5THz, which is the center frequency of most THz sources. The customized nanotip can achieve 40 nm $(\lambda / 15000)$ spatial resolution while maintaining an excellent scattering efficiency. This nanotip design method doesn’t depend on the homogenization commercial AFM tips, providing an approach for customized nanotip design of THz wave scattering near field microscopy.
\end{abstract}

\section{Introduction}
{
Terahertz (THz) radiation locates an important position in the electromagnetic spectrum. THz spectroscopy has been widely applied to resonantly probe collective charge, spin, and lattice oscillations in solids, the rotations of small polar molecules, and the structural vibrations of large biomolecules. THz imaging has become the distinctive detection technique in biological tissue samples and new materials. However, the spatial resolution with conventional optical methods is constrained by the diffraction limit, which obstructs the application and development of THz imaging. Scattering scanning near field microscopy (s-SNOM) is a promising scanning probe technique, which conveys near-field dielectric information by nano tips scattering and is detected by conventional far-field detectors \cite{huber2008terahertz,huth2012nano}. The surface charge density of the tip apex is polarized by the incident field, and the surface charge is concentrated to the nanoscale by the tip apex \cite{liewald2018all,mcleod2014model}. When the nanotip-sample distance reduces to the near-field region, the tip scattering field is influenced by the local tip-sample system’s dielectric properties. The nanotip-sample system scattering field can transfer and record at the far field. The dielectric response and surface topography can be recorded simultaneously \cite{huth2013resonant,mastel2017terahertz,gomila2014finite,chen2004identification}. 

In the SNOM system, the spatial resolution and scattering efficiency are predominantly determined by the tip geometry, as the nanoscale tips provide the basic near-field environment \cite{bouhelier2006field}. Besides, the dielectric response of the sample surface and incident field intensity and polarization also affect the scattering efficiency. In THz frequency, the low THz source power and the long wavelength limit the development of THz s-SNOM. High power CW THz source (QCL laser, CO2 laser) has been used to build THz s-SNOM system, however, the CW THz sources are not equipped time resolution ability inevitably. The photoconductive antenna is an appropriate pulsed THz source with time resolution ability and its high repetition is very suitable for the demodulation of near-field signals. However, its low pulse energy can’t support near-field THz nonlinear experiments. High pulse energy THz sources like two-color laser filaments, lithium niobates, and free electron laser are all restricted by the low repetition, which is not suitable for the near field signal demodulation (low signal-to-noise ratio). Therefore, increasing the near field scattering signal intensity while maintaining spatial resolution is a promising method to introduce high-energy THz source into s-SNOM.

Systematic studies and optimization of the scattering probe always focus on the single frequency THz field and narrow band THz source, as most commercial THz-SNOM equipment is based on CW THz source (QCL laser, CO2 laser) or narrow band THz source (photoconductive antenna). This letter comprehensive characterizes the performance improvement for the near-field spatial resolution and scattering efficiency induced by nanotips’ geometry and incident field. The theoretical model is numerical full wave simulation based on solving Maxwell’s equations in the frequency domain. The spatial resolution is dominated by the tip apex radius, tip cone angle and tip-sample distance. The near field scattering efficiency is dominated by the tip length and incident field polarization and incident angle. Besides, this letter optimizes the tip’s geometry for a strong pulsed THz source based on two-color femtosecond laser filament, as it’s a high pulse power, high efficiency, and broadband THz source. The THz energy is mainly concentrated around 0.5THz, so this letter optimizes the tip geometry with 0.5THz to simulate the real frequency spectrum based on two-color laser filament. Finally, this letter provides an achievable optimization nanotip design scheme for the THz-SNOM system based on high power pulsed THz source, which extends the high-pulsed energy THz source application in ultra-resolution imaging.

}

\section{Theoretical model}
{
Geometric considerations of scattering tips are essential for a proper description of near-field scattering signals. A typical tip-sample dipole model has been used to analyze the focusing spots diameter and the focusing intensity of the scattering tips. However, the tip is simplified as an ideal nanosphere in the tip-sample dipole model, so the tip length and tip cone angle can't be accurately characterized. It indicates that the tip-sample dipole model can’t describe the integral electromagnetic response of the tips. 

An alternative method is the numerical simulations based on solving Maxwell’s equations in the frequency called Full Wave Numerical Simulations \cite{maissen2019probes,chen2020thz,mastel2018understanding}. The electric dipole moment of charge distribution is defined by the integral expression, in which the tip-sample system is confined to a compact region of space:

\begin{equation}
 \mathbf{p}(t)=\int d^3 x^{\prime} \mathbf{x}^{\prime} \rho\left(\mathbf{x}^{\prime}, t\right)
\end{equation}

In equation (1), $x^{\prime}$ and $\mathbf{x}^{\prime}$ are the boundary conditions of Green function G ($x^{\prime}$,$\mathbf{x}^{\prime}$) which describes the electromagnetic response of the nanotip scattering and sample surface scattering. $\rho\left(\mathbf{x}^{\prime}, t\right)$ determines the charge density of the nanotip apex surface. Charge conservation relates the charge density to the current density by the continuity equation:

\begin{equation}
\nabla \cdot \mathbf{J}+\frac{\partial \rho}{\partial t}=0
\end{equation}

$\mathbf{J}$ is the current density. The following relation between the electric dipole moment and the current density follows:

\begin{equation}
    \frac{d}{d t} \mathbf{p}(t)=\int d^3 x^{\prime} \mathbf{J}\left(\mathbf{x}^{\prime}, t\right)
\end{equation}

\begin{equation}
    \mathbf{J}=\left|\begin{array}{ccc}
\mathbf{i} & \mathbf{j} & \mathbf{k} \\
\frac{\partial}{\partial x} & \frac{\partial}{\partial y} & \frac{\partial}{\partial z} \\
H_x & H_y & H_z
\end{array}\right|-(-i \omega \mathbf{D})
\end{equation}

As it is shown in equation (3), the electric dipole moment is proportional to the current density. The tip-scattered electric field $E_{\mathrm{sca}}$ is proportional to the complex-valued dipole moment $\mathbf{p}(t)$, calculated numerically according to:

\begin{equation}
    E_{\mathrm{sca}} \propto \mathbf{p}(t)=\int \sigma(\mathbf{r}) \mathbf{r} d \mathbf{r}
\end{equation}

\begin{equation}
    \sigma \propto \mathbf{E} \cdot \mathbf{n}
\end{equation}

$\sigma(r)$ is the surface charge density, $\mathbf{r}$ is the ideal nanotip's apex radius, $\mathbf{E}$ is the electric field vector and $\mathbf{n}$ is the outward normal to the tip surface. 

\begin{figure}[h!]
\centering\includegraphics[width=7cm]{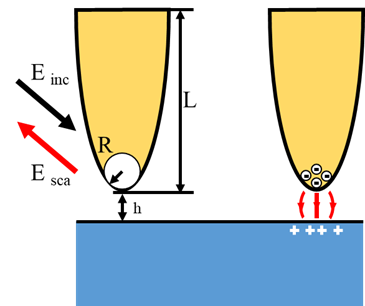}
\caption{Sketch of Numerical Full-wave simulations and near field interaction. $E_{\mathrm{inc}}$ is the incidence wave, $E_{\mathrm{sca}}$ is the near-field scattering signal. }
\end{figure}

Fig.1 shows the sketch of the near scattering system. In this paper, the tips are modeled as a conical frustum and a semi-spherical apex. The tip length is 20, 40, 80, and 160 $\mu \mathrm{m}$, the tip apex radius is 25, 50, 200, and 400 nm, the tip-sample distance is 5, 10, 15, and 20 nm, and the cone angle is 10° and 20°. The tip-sample system is illuminated by a 0.5THz linear polarization field. The Au nanotips and the Au surface are described by the Drude model with plasma frequency $\omega_{\mathrm{pl}}$ = $2 \pi \times 2100 \mathrm{THz}$, and the collision frequency $\gamma=2 \pi \times 50 \mathrm{THz}$ \cite{klein2006second}. 
}

\section{Design method}
{
Preliminarily, design the dimensions of the nanotips (cone angle, length, and apex diameter) based on the conventional constraints of scanning microscopy. In scanning microscopy, nanotips’ cone angle is usually < 40°, that because the excessive cone angle would cause loss of the sample surface morphology information when scanning the sample surface. And nanotips’ length is usually below 200 $\mu \mathrm{m}$, which meets incident field wavelength design requirements. Apex diameter (apex curvature) similarly determines the scanning accuracy, which needs to be considered with scattering efficiency.

\begin{figure}[htbp]
\centering\includegraphics[width=13cm]{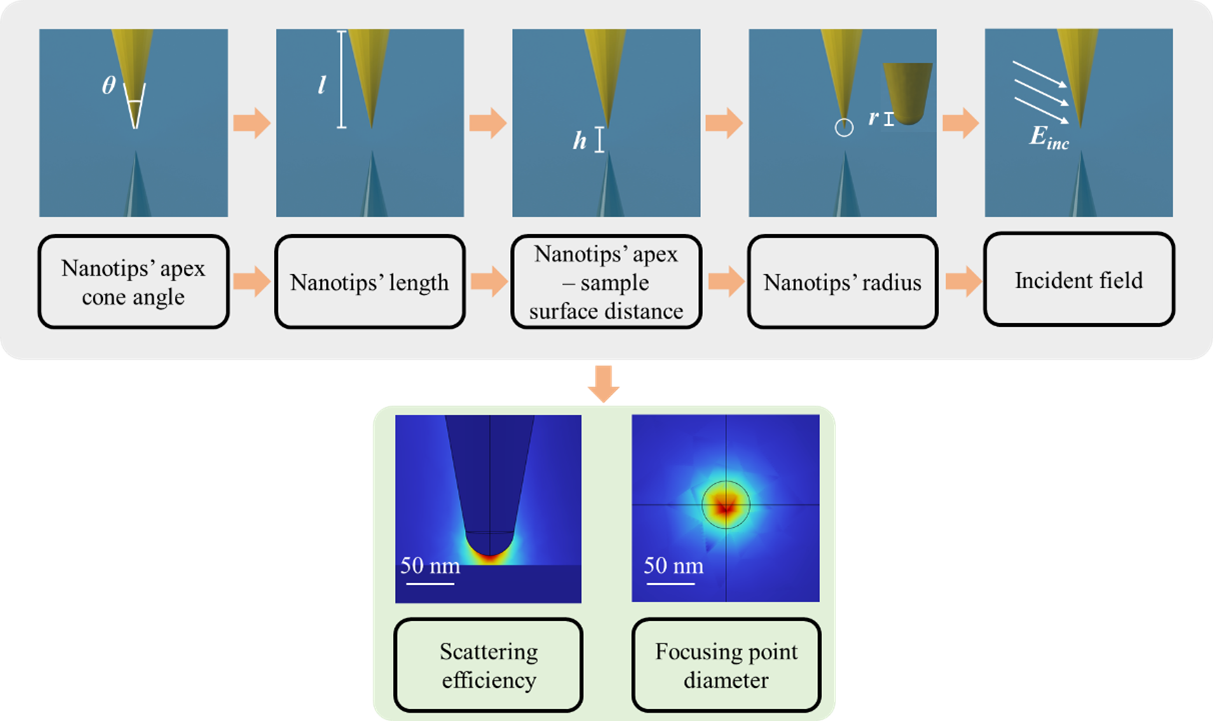}
\caption{The design process of the nanotip of THz s-SNOM.}
\end{figure}

On the basis of conventional constraints of scanning microscopy, this paper focuses on purposing a nanotip design method to improve the THz s-SNOM spatial resolution and scattering efficiency by solving Maxwell’s equations in the frequency (Full Wave Numerical Simulations). As most of the strong THz sources’ center wavelength is ~0.5THz, the design process and simulation are pointing at 0.5THz. 

Fig.2 demonstrates the design process details of the nanotips used in THz s-SNOM. Nanotips’ geometric parameters design is the preliminary work. The geometric parameters of the nanotips affect the THz focusing efficiency at the nanotips’ apex. After confirming the geometric parameters of the probe, it is necessary to comprehensively consider the influence of the probe and sample surface system on the scattering signal. THz s-SNOM makes use of evanescent fields that exist only near the surface of the sample. This field carries high-frequency spatial information about the sample surface and has intensities that drop exponentially with distance from the sample surface. Finally, the optimal incident light polarization conditions need to be calculated based on the nanotips with ideal parameters. As the scattering signal is also affected by the incident field. The nanotips can be seen as a near-field antenna, which will be influenced by the incident field properties. 
}

\section{Result and discussion }
{
Atomic force microscopes and scanning tunneling microscopes are commonly used to provide near-field environments for the tip-sample system. However, conventional scanning tips’ length locates in the subwavelength scale compared with the THz wavelength. Therefore, although the scattering-type scanning near-field optical microscope (s-SNOM) is essentially independent of the incident wavelength, optimizing the coupling and scattering efficiency are still great challenges. 

The tip is essential for an antenna to acquire the THz signal and confine the electromagnetic field below the tip apex. Near-field microscope studies in infrared frequencies have shown that by designing a tip to act as a resonant antenna, the scattering efficiency can be significantly improved. The resonant antenna fundamental principle is a metallic scattering support resonant mode when it is of a similar length to the incident wavelength. In THz frequencies, most conventional tips tend to be much longer than the THz wavelength, especially corresponding to high pulsed energy THz sources (two-color laser filament), in which most energy is located in the low frequencies range. 

An approximate relationship between the tip length and the resonant wavelength of its fundamental dipole resonance is given by the commonly used ideal half-wavelength dipole model. However, the ideal resonant antenna model does not account for variation geometry in the nanotips, as most nanotips’ geometry is approximately expressed as cone or pyramid type, instead of a cylindrical antenna. Besides, in the near-field scattering system, the scattering tip and the sample surface need to be regarded as a complete dielectric environment. Method of images is used to describe the nanotip sample near field scattering system. Nanotip is simplified to an electric point dipole with dipole moment $p$ above the sample surface (conducting plane), and the image of the electric dipole moment with equal magnitude and direction rotated azimuthally by $\pi$. The dipole moment $p$ is related to the nanotip geometry, tip-sample distance, and incident power. Finally, in this letter, the dipole antenna theory and method of the image are used to analyze and design the nanotip geometry. The simulation is based on solving the Maxwell equation in the frequency.  

\begin{figure}[htbp]
\centering\includegraphics[width=9cm]{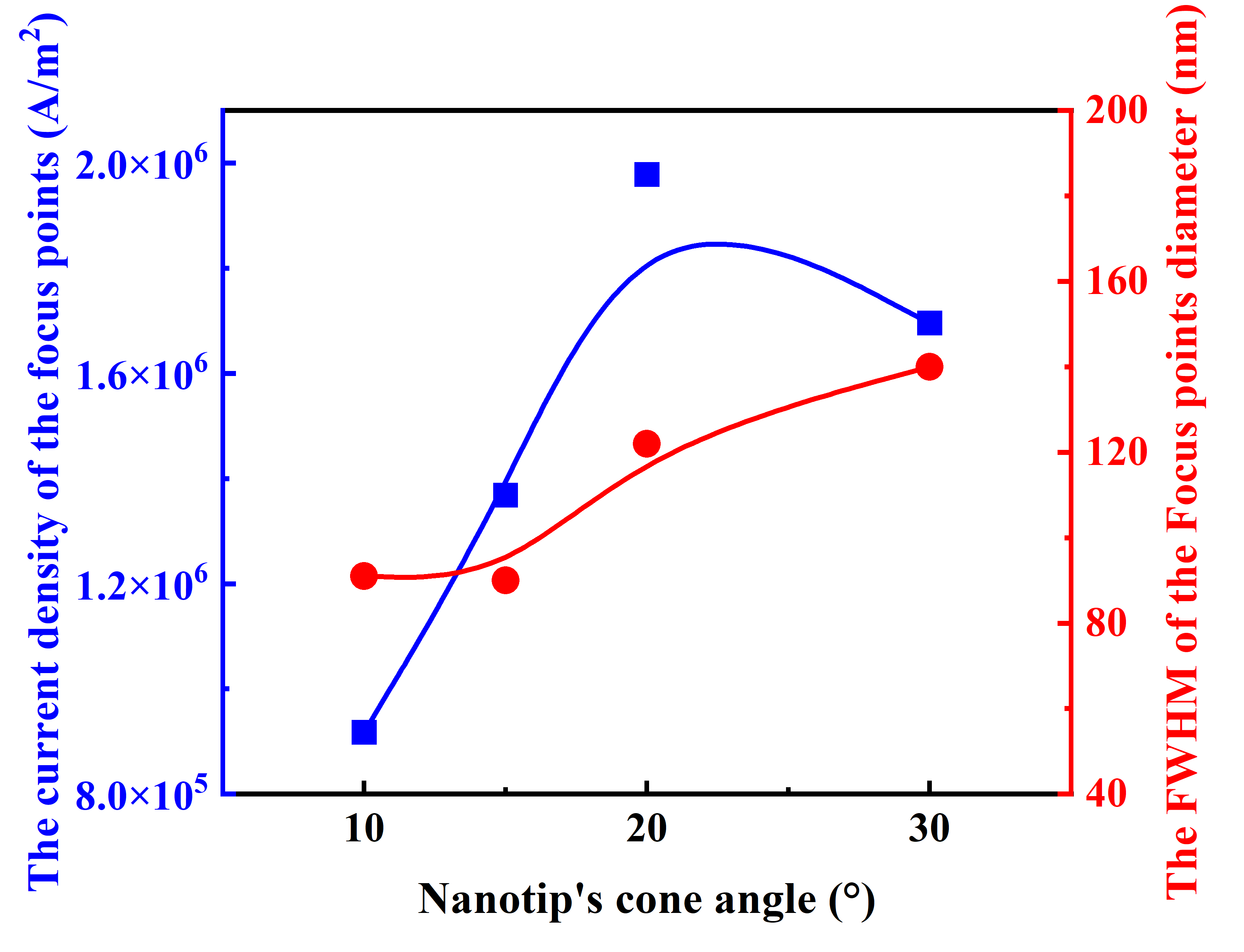}
\caption{The evolution trend of the scattering efficiency and the FWHM of the focus diameter are influenced by Nanotip’s cone angle.}
\end{figure}

In the scattering microscopy system, the nanotip’s cone angle determines the scanning accuracy of the characterization of sample surface morphology, as a wide-angle cause the nanotip cannot achieve precise scanning of the depth information. Similarly, a wide-angle is not suitable for the THz-SNOM system \cite{cory1998electric}. Considering the behavior of the enhancement of the scattering efficiency influenced by the cone angle is necessary, that because different lengths of resonance nanotips correspond to different optimal cone angles. In Fig.3, the scattering efficiency evolution is nonmonotonic, and the optimal cone angle for which the maximal localize enhancement is 20° for the nanotip. Meanwhile, the spatial resolution is also influenced by the cone angle. Different from the evolution of the scattering efficiency, the focus point diameter increases gradually as the cone angle increase from 10° to 30°. The quantity treatment of this phenomenon can be described by the full wave simulation, while the point-like dipole approximation is inadequate to quantitatively describe the enhancement near the apex. More specifically, the near field is thus practically determined by the surface current density near the apex. The evolution of the cone angle fundamentally changes the surface density near the apex. Apart from the optimal cone angle, nanotip length exist an optimal value to support the maximum electric field enhancement near the apex.

Thin linear conductors of length $l$ are in fact resonant at any integer multiple of a half-wavelength. In the nanotip sample system, the half wave dipole antenna consists of the scattering tip and sample surface (image dipole moment). As it is shown in the Fig.4, cone nanotips with cone angle 20° and apex radius 25 nm is simulated for cone lengths $L$ = 20, 40, 80 and 160 $\mu \mathrm{m}$. The incident field frequency is 0.5 THz. The blue line shows the current density, calculated by the full wave simulation, of the focus points in the sample surface, which indicates the optimizing nanotip length is 160 $\mu \mathrm{m}$ (a quarter of the incident wavelength). The simulation result is according to the analysis mentioned before. The nanotip sample system consists of an electric magnetic environment ensemble, so the ideal nanotip length is quarter of the incident wavelength. However, the spatial resolution doesn’t mainly determine by the nanotips’ length. In the red line, the Full width at half maximum (FWHM) doesn’t change significantly with the changes in the nanotips’ length. The nanotips’ length determines the near field couple and scattering efficiency, while the focus resolution is affected by other fine structures of nanotips’ apex. 

\begin{figure}[htbp]
\centering\includegraphics[width=9cm]{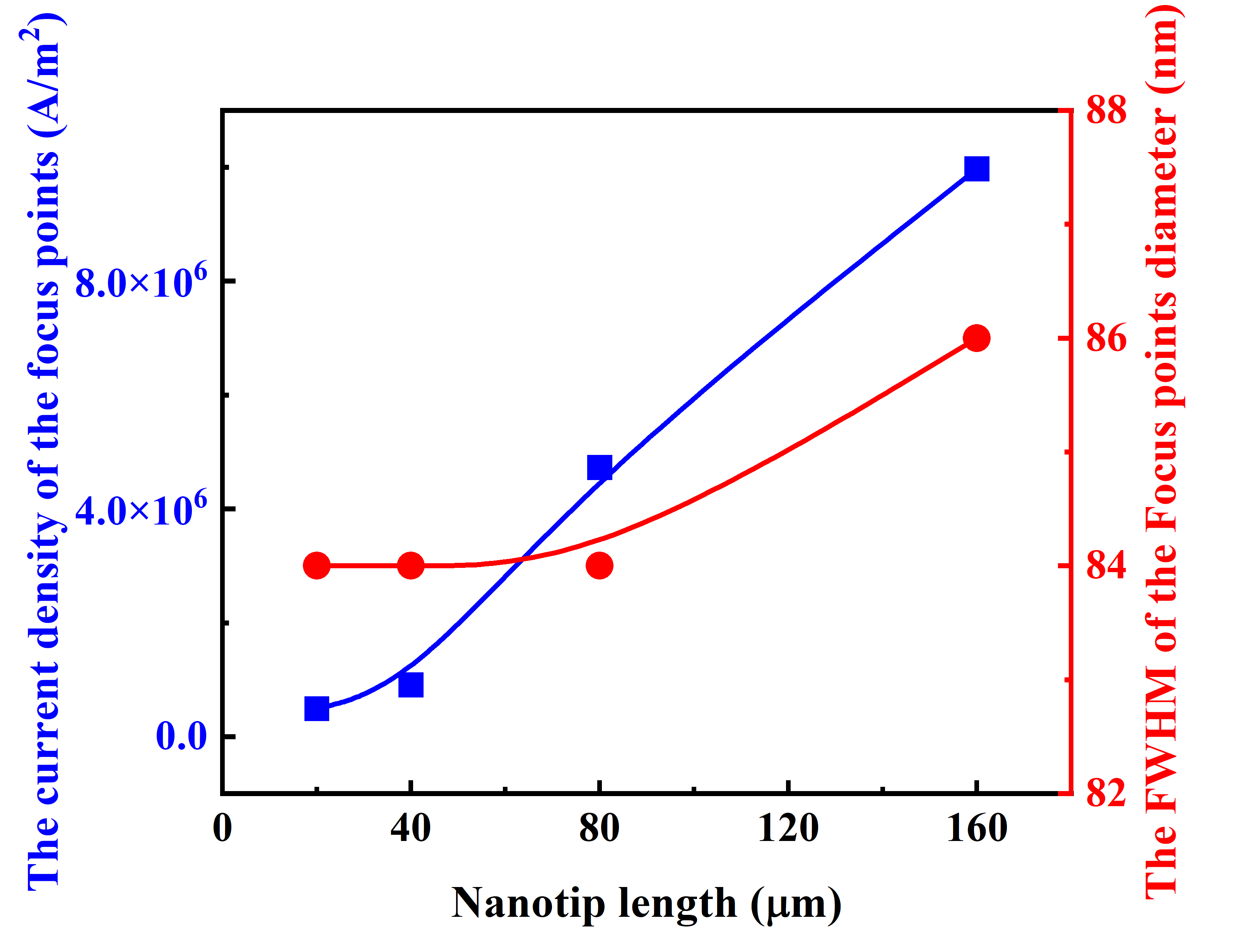}
\caption{The evolution trend of the scattering efficiency and the FWHM of the focus diameter are influenced by Nanotip’s length.}
\end{figure}

Fig.5 shows the current density and focus point diameter of the focus points for several nanotips that are statically placed at a distance above the sample. Nanotips’ parameters are as follows: cone angle 20°, length 160 $\mu \mathrm{m}$, and the nanotips-sample distance h = 5, 10, 15, and 20 nm. From the full wave simulation, the near field interaction intensity is rapidly increased for the decrease of the nanotip-sample distance. This obvious near-field signal enhancement can be attributed to the increase in the dipole moment. The near-field interaction above the sample surface can be represented as a superposition of the fields of nanotip dipole moment and the image dipole moment (below the sample surface). Nanotip-sample dielectric system’s effective polarizability according to:

\begin{equation} \mathbf{p}=\alpha\left[\mathbf{E}_o+\mathbf{E}_{\mathrm{image}}\right] \equiv \alpha_{\mathrm{eff}} \mathbf{E}_o
\end{equation}

$\mathbf{p}$ is the dipole moment, $\alpha_{\mathrm{eff}}$ is the effective polarizability, and $\mathbf{E}_o$ is the incident field. The effective polarizability along the nanotip direction is shown as follows:

\begin{equation} \alpha_{\mathrm{eff}}=\frac{\alpha[1-\beta]}{1-\alpha \beta /\left[32 \pi \varepsilon_o(r+h)^3\right]}
\end{equation}

 $\alpha$ is the nanotips’ polarization response of the incidence wave. $\beta$ is the quasi-static Fresnel-reflection coefficient. From the equation (7) and (8), the scattering efficiency depends on the nanotip-sample distance. Meanwhile, the increased near-field interaction changes the electric charges distribution below the nanotip. When the nanotip-sample distance decrease, the local surface charge density is drastically increased in this region, leading to stronger confinement, which reduces the focus point diameter, of the localization field. That is spatially confined by the nanotip apex. This is an expression of nanotip apex inducing near-field scattering spatial localization. 

\begin{figure}[htbp]
\centering\includegraphics[width=9cm]{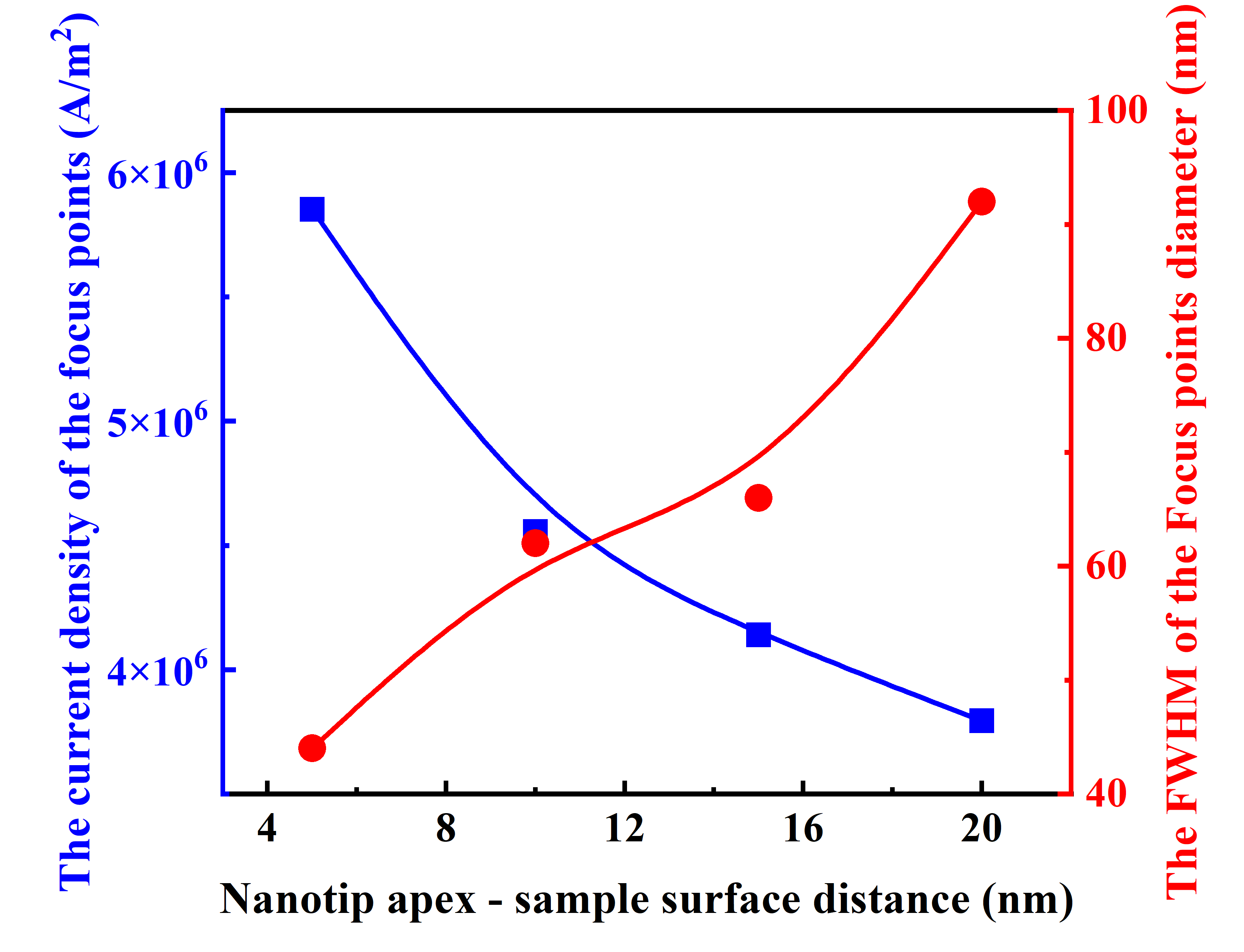}
\caption{The evolution trend of the scattering efficiency and the FWHM of the focus diameter are influenced by Nanotip apex-sample surface distance.}
\end{figure}

From equation (8), the apex radius is another expression of the nanotip apex inducing near-field spatial localization. Systematically study the influence of the nanotips’ apex radius on the scattering efficiency and the spatial resolution is shown in Fig.6. According to the optimized geometry parameters of the nanotips, the cone angle is 20°, the nanotips’ length is 160 $\mu \mathrm{m}$. As it is shown in Fig.6, the scattering efficiency shows gradual attenuation, as the nanotip apex radius decrease from 400 nm to 50 nm. However, the scattering efficiency rapidly decreases by 1.3 times, as the nanotips’ apex radius decreases from 50 nm to 25 nm. Interestingly, the focus point diameter Evolution trend is different of scattering efficiency. The focus point diameter decreases rapidly as the nanotip apex radius decreases from 400 nm to 50 nm. The focus point diameters from nanotip apex 25 nm and 50 nm aren’t significantly different. Therefore, a smaller nanotip apex radius is instrumental in improving spatial resolution. However, this evolution trend isn’t obvious when the nanotip apex radius is below 50 nm, more than that, the scattering efficiency rapidly decreases simultaneously. This conclusion is crucial for designing the scattering nanotip’s geometry, that because it is necessary to improve the signal-to-noise ratio while ensuring the scattering efficiency has experimental significance.

\begin{figure}[htbp]
\centering\includegraphics[width=9cm]{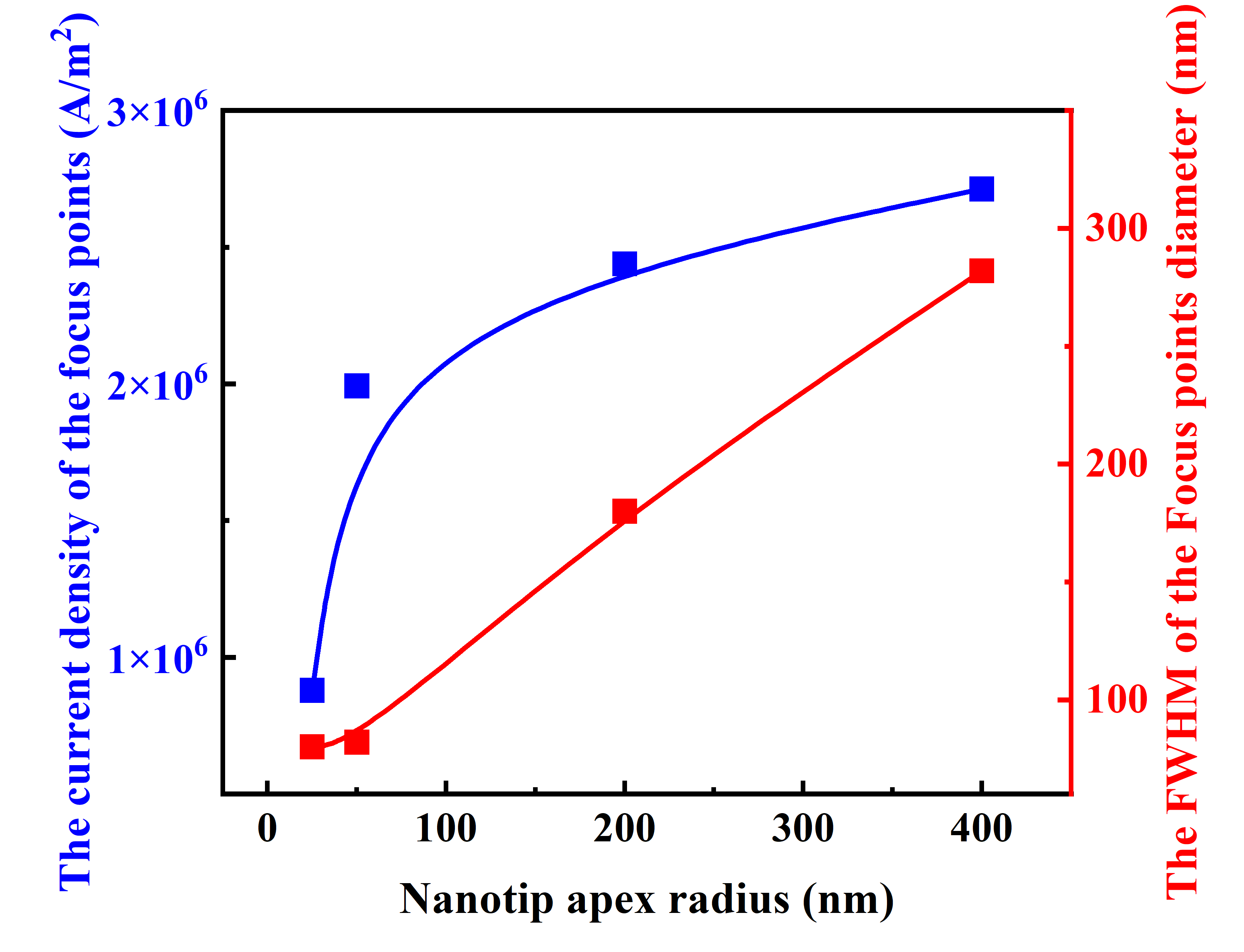}
\caption{The evolution trend of the scattering efficiency and the FWHM of the focus diameter are influenced by Nanotip apex radius. }
\end{figure}
 
Apart from the scattering nanotip’s geometry, the incident field polarization is also influencing the scattering efficiency. The relationship between the scattering field $I$ and the incident field polarization can be simplified present as:

\begin{equation}
    I(\beta) \propto \sin ^2(\pi / 2-\beta)
\end{equation}

$\beta$ is the included angle of the nanotip’s direction and the polarization\cite{aigouy1999polarization}. From equation (9), the near field polarization reaches a maximum when the electric field vector of the incident wave has parallel to the nanotip. As it is shown in Fig. 6, the current density of the focus point scattered by the nanotips is 2 orders of magnitude higher when the incident field polarization is parallel to the nanotip direction than when it is vertical to the nanotip direction. Although the incident field polarization influences the scattering efficiency, the spatial resolution is independent of the incident field polarization as it is shown in the Fig.7 red line. 

\begin{figure}[htbp]
\centering\includegraphics[width=9cm]{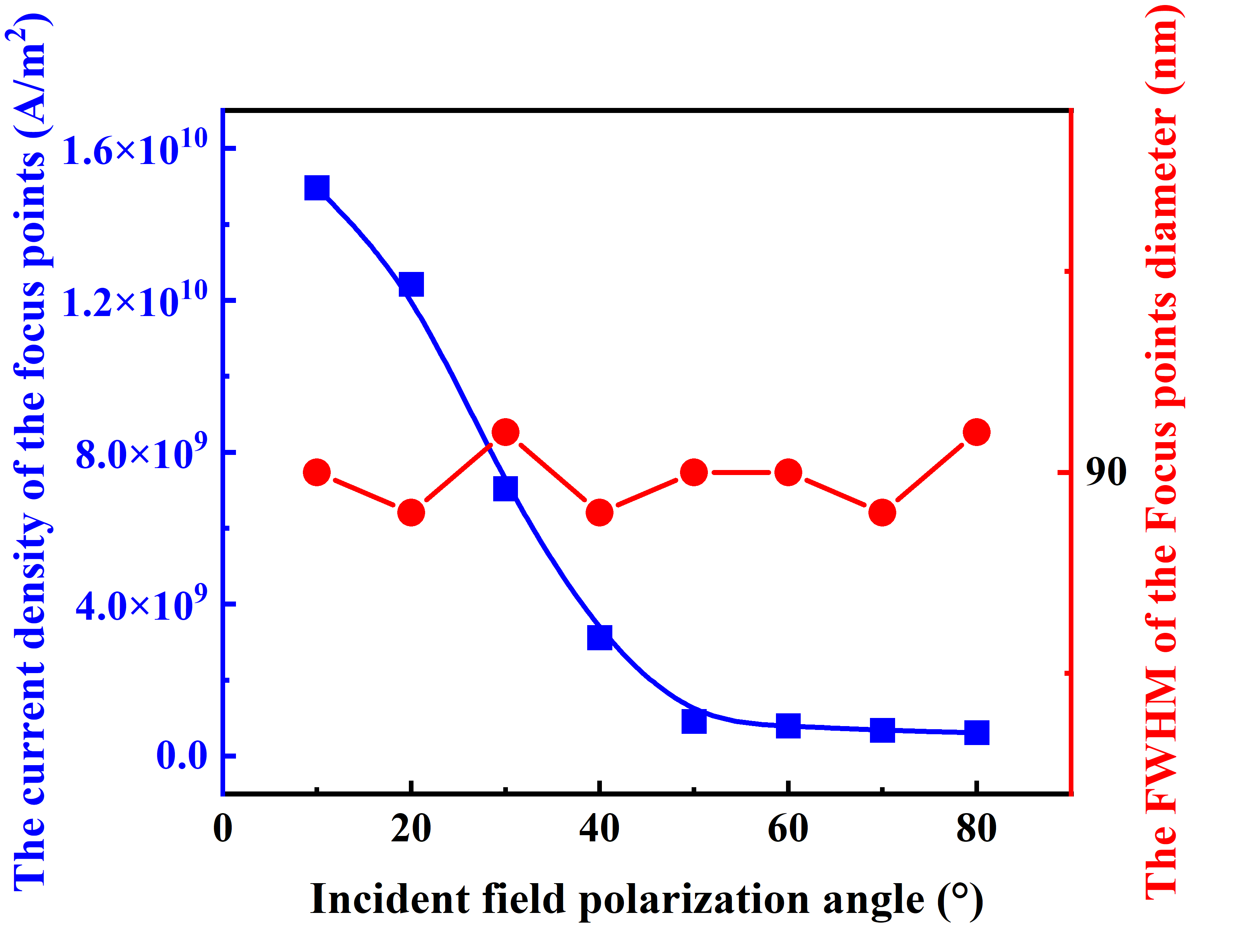}
\caption{The evolution trend of the scattering efficiency and the FWHM of the focus diameter are influenced by incident field polarization status. }
\end{figure}

\begin{figure}[htbp]
\centering\includegraphics[width=13cm]{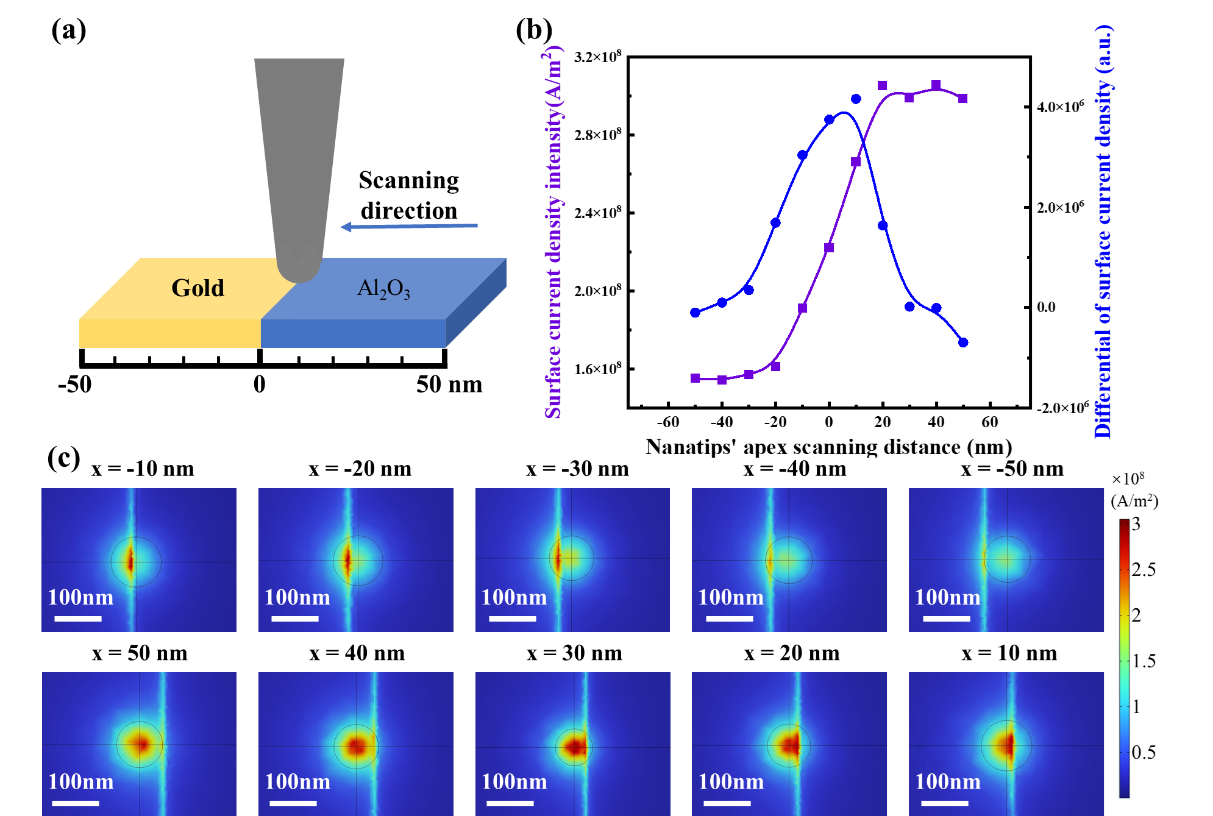}
\caption{Numerical simulation of THz near filed line scanning. (a) Sketch of the geometry. A tip with apex radius $r$ = 50 nm and length 160 $\mu \mathrm{m}$ is placed above a sample consisting of Gold on the left ($x$ < 0 nm) and $\mathrm{Al}_2 \mathrm{O}_3$ on the right ($x$ > 0 nm) side. (b) Intensity evolution curve of THz near field scattering signal. (c) Electric near-field distribution below the tip apex for different tip positions. }
\end{figure}

In the demonstration above, the optimal parameters of the nanotips’ geometry are comprehensive discussion. Considering the scattering signal intensity (current density of the focus points) and the FHWM of the focus point, a conical tip of 160 $\mu \mathrm{m}$ length and apex radius $r$ = 50 nm is placed above a sample modeled by Gold on the left side, and $\mathrm{Al}_2 \mathrm{O}_3$ on the right side. The material boundary at $x$ = 0 nm. The polarization of the input wave is parallel along the nanotips at 0.5 THz at an illumination angle of 60° relative to the tip axis. Fig.8 (a) shows the scanning process of the simulation. The purple curve in Fig.8 (b) shows the result of surface current density evolutionary trends at the nanotips’ apex, which represents symmetry essentially on the transition boundary of two kinds of materials. This symmetry can be explained by the optimization of the nanotips’ geometry. The scanning spatial solution, depending on the nanotips’ geometry, which is described by the differential curve of the surface current density.

Fig.8 (c) shows the transverse scanning process continuously. The near-field scattering signal shows significant changes when nanotips cross two materials’ boundaries. Similar to equation (8), the difference in the different scattering boundaries can be attributed to the variation of the local-dielectric function of the sample based on the method of image charges. In the effective polarizability equation (8), the local dielectric function is included in the quasi-static Fresnel-reflection coefficient $\beta$. $\varepsilon_{sample}$ is the dielectric constant of the materials. 

\begin{equation}
    \beta=\frac{\varepsilon_{\text {sample }}-1}{\varepsilon_{\text {sample }}+1}
\end{equation}

The scattering efficiency of the nanotips depends on the dipole (and the image dipole) moment, which can be described as effective polarizability. The scattering signal is influenced by nanotip geometry, nanotip sample distance, local dielectric properties of materials, and the polarization of the incidence wave. The near field scattering signal characterizes dielectric properties of the local materials by scanning the structure of the material and controlling the other factors unchanged. In Table 1, the THz s-SNOM performance of this paper is compared with several representativeness THz s-SNOM systems. To that end, the nanotips’ geometry, incident field wavelength, and resolution are chosen as testing standards. Although the nanotips’ apex diameter in this paper is bigger than \cite{pogna2021mapping,maissen2019probes,plankl2021subcycle,guo2021near}, we still improve the resolution to 40 nm with 0.5THz ($\lambda$ = 600 $\mu \mathrm{m}$). The optimization of the nanotips’ geometry can noticeably improve the spatial resolution, especially for specific incident wavelengths. This method can provide the optimized nanotip parameters and assist the search for the balance parameters between the scattering efficiency and spatial resolution. The nanotips design process and method mentioned in this paper provide a new method for improving the spatial resolution of THz s-SNOM.

\begin{table}[]
\begin{tabular}{ccccccc}
\hline
\textbf{\begin{tabular}[c]{@{}c@{}}Parameter\\ comparison\end{tabular}} & \multicolumn{6}{c}{\textbf{Comparison of Multiple Design Schemes}} \\ \hline
\begin{tabular}[c]{@{}c@{}}Nanotip's apex\\ (nm)\end{tabular}           & 10\cite{pogna2021mapping}        & 6\cite{maissen2019probes}       & 20\cite{plankl2021subcycle}       & 20\cite{guo2021near}       & 50(this paper)        & 100\cite{chen2020thz}       \\
\begin{tabular}[c]{@{}c@{}}Incident field\\ (THz)\end{tabular}          & 3         & 2.52     & 1.3      & 0.5      & 0.5       & 0.15      \\
Resolution (nm)                                                         & 30        & 15       & 50       & 500      & 40        & 200       \\
Resolution (1/$\lambda$)                                   & 1/3333    & 1/8000   & 1/4600   & 1/1200   & 1/15000   & 1/10000   \\ \hline
\end{tabular}
\end{table}

}

\section{conclusion}
{
In conclusion, the full wave simulation is used to describe the ensemble dielectric response of the scattering probe in THz frequency. The nanotip’s geometry (cone angle, length, and apex radius), incident field, and local material dielectric constant are comprehensively considered to obtain the optimal parameters. The nanotip’s cone angle influences the superficial area and the dielectric permittivity near the apex, thus the influence of the cone angle on scattering efficiency is not monotonic. The nanotip’s length design is according to dipole moment and resonant antenna analysis, which can enhance the scattering efficiency 3 times. The spatial resolution mainly depends on the nanotip’s apex radius and distance between the apex and sample surface. The ultimate resolution reaches 40nm with a 50 nm radius nanotip. The scattering efficiency is also depending on the polarization of the incident field. When the incident field polarization is parallel to the nanotip's direction, the scattering efficiency can enhance 2 orders. Finally, this letter presents a feasible model to extract the optimal parameters of the THz near field scattering system. Especially, this letter optimization focuses on most THz sources (~0.5THz), which provides a customization method to introduce this THz source into the THz-SNOM system.
}
\\
\\
\begin{backmatter}
\bmsection{Acknowledgments}

The authors would like to thank the funding support from National Natural Science Foundation of China (12061131010, 12074198), Russian Science Foundation (21-49-00023), Fundamental Research Funds for the Central Universities (63223052).

\bmsection{Disclosures}The authors declare that they have no known competing financial interests or personal relationships that could have appeared to influence the work reported in this paper.

\bmsection{Data availability}Data underlying the results presented in this paper are not publicly available at this time but may be obtained from the authors upon reasonable request.

\end{backmatter}



\end{document}